\documentclass[pra,12pt,preprint,nofootinbib,tightenlines]{revtex4} 
\usepackage{amsmath}
\begin{document}
\def\vs{\vskip .2in} 

\title{A Measure of Classicality\footnote{\sl The core of this paper is a  section of the authors' paper {\it Strong Decoherence \cite{GH95} }  modestly edited by the junior (second)  author.  The  section is concerned with the general problem of a measure for classicality and a specific proposal for that measure.  It is  is largely self-contained and can be read separately from the rest with the help of the modest edits by the junior author. Neither the text or the references have been updated except for a few ``to be published'' references. Sections III through V have been added by the junior author.}}  

\author{James B.~Hartle}
\affiliation{Department of Physics, University of California,Santa Barbara, CA 93106-9530,  \\ and
Santa Fe Institute, Santa Fe, NM 87501}
\email{hartle@physics.ucsb.edu}

\author{Murray Gell-Mann{\large$\dagger$}}
\affiliation{
Santa Fe Institute, 1399 Hyde Park Rd
Santa Fe, NM 87501,\\
Los Alamos National Laboratory, MS-B210
Los Alamos, NM 87545 \\ and
Department of Physics and Astronomy, University of New Mexico Albuquerque, NM 87131}


\begin{abstract}
A striking feature of our fundamentally indeterministic quantum universe is its quasiclassical realm --- the wide range of time place and scale in which the deterministic laws of classical physics hold.  Our  quasiclassical realm
is an emergent feature of the fundamental theories of our universe's quantum state and dynamics. There are many types of quasiclassical realms our Universe could exhibit characterized by different variables, different levels of coarse-graining, different locations in spacetime, different classical physics, and different levels of classicality.We propose a measure of classicality for quasiclassical realms, We speculate on the observable consequences of different levels of classicality especially for information gathering and utilizing systems (IGUSes) such ourselves as observers of the Universe.
\end{abstract}




\maketitle

\section{Introduction---Our Quasiclassical Realm} 
\label{our-qclm}
The most striking observable feature of our indeterministic quantum universe is the wide range of time, place, and scale on which the deterministic laws of classical physics hold to an excellent approximation defining what we can call our {\it quasiclassical realm}.
The regularities that characterize a   quasiclassical realm are described by the familiar classical equations for particles, bulk matter, and fields, together with the Einstein equation governing the regularities of classical spacetime geometry. Our observations  suggest that  our quasiclassical realm extends over the  patch of classical spacetime that is visible to us through large scale observations.\footnote{In simplified models like the FLRW ones a quasiclassical realm is assumed to  extend much further sometimes from just after the big bang but this requires neglecting  phenomena like eternal inflation. see, e.g  \cite{EIinout}}.  What is the origin of this predictable quasiclassical realm in a quantum universe characterized by indeterminacy and distributed probabilities?  

Our quasiclassical realm is is an  observed feature or our Universe  on a par  with such observed features as its expanding  classical spacetime, its approximate homogeneity and isotropy on large distance scales,  the fluctuations away from these symmetries that we see in the cosmic background radiation and the statistics of  galaxies, and  in the primordial abundance of the elements.

In various papers developing decoherent (or consistent) histories quantum mechanics (e,g, \cite{GH90a,GH90b}) the authors have defined a quasiclassical realm as a decohering  set of alternative coarse-grained  histories of the Universe for which the probabilities are high for histories that exhibit  correlations in time governed by deterministic classical laws like the Einstein equation,  the Navier-Stokes equation, or Maxwell's equations, etc. \cite{Har11}.

A typical history in a quasiclassical realm will describe long stretches in time of classical behavior interrupted by non-classical events such as quantum fluctuations, quantum transitions,  and quantum measurements.  It is  for this reason that we refer to {\bf quasi}classical realms rather than  classical realms.





Quantum mechanics, along with the correct theory of the
elementary particles (represented by the Hamiltonian $H$ and the
correct initial condition in the universe (represented by the
state vector $|\Psi\rangle$), presumably exhibits a great many
essentially different  decohering realms, but only some of those
are quasiclassical.  For the quasiclassical realms to be
viewed as an emergent feature of ($H$, $|\Psi\rangle$), and quantum
mechanics, a good technical definition of classicality is required. (One
can then go on to investigate whether the theory exhibits many
essentially inequivalent quasiclassical realms or whether the
ours is nearly unique.)

\section{A Measure of Classicality} 
\label{measure}

In earlier papers, \cite{GH90a,GH93a,GH90b} 
 we have made a number of suggestions about the
definition of classicality and it is appropriate to continue that
discussion here.  It is clear that from those earlier discussions that
classicality must be related in some way to a kind of entropy for
alternative coarse-grained histories.  We must therefore begin with an
abstract characterization of entropy and then investigate the
application to histories. An entropy $S$ is always associated with a
coarse graining, since a perfectly fine-grained version of entropy in
statistical mechanics would be conserved instead of tending to increase
with time.   Classically, if  the set of all fine-grained alternatives are
designated by $\{r\}$, with probabilities $p_r$ summing to one, that
fine-grained version of entropy would be
\begin{equation}
S_{f-g} = -\sum\limits_r p_r \log\, p_r\ ,
\label{fiveone}
\end{equation}
where $\log$  means  $\log_2$ and where, for convenience, we
have put Boltzmann's constant $k$  times $ {\rm log}_e 2$
equal to unity. A true,
coarse-grained entropy has the form
\begin{equation}
S \equiv   -\sum\limits_r \tilde p_r \log \tilde p_r\ ,
\label{fivetwo}
\end{equation}
where the probabilities $\tilde p_r$ are coarse-grained averages of the
$\{p_r\}$. A coarse graining $p_r\to\tilde p_r$ must have
certain properties (see \cite{GL96} for more details):
\def\tp{{\tilde p}}
\begin{subequations}
\label{fivethree}
\begin{eqnarray}
{\rm 1)} &\qquad\qquad&{\rm the}\ \{\tilde p_r\}\ {\rm are\ probabilities}
\ ,\label{fivethreea}\\
{\rm 2)} &\qquad\qquad& \skew6\tilde\tp_r = \tilde p_r\ ,
\label{fivethreeb}\\
{\rm 3)} &\qquad\qquad& -\sum\limits_r p_r\log\, \tilde p_r =
-\sum\limits_r \tilde p_r \log\, \tilde p_r\ .
\label{fivethreec}
\end{eqnarray}
\end{subequations}
These properties are not surprising for  an averaging procedure. The
significance of the last one is easily seen if we make use of the well known
fact
that for any two sets of probabilities $\{p_r\}$ and
$\{p^\prime_r\}$ we have
\begin{equation}
-\sum\limits_r p_r \log\, p_r \leq - \sum\limits_r p_r \log
\, p^\prime_r\ .
\label{fivefour}
\end{equation}
Putting $p^\prime_r = \tilde p_r$ for each $r$ and using
(1)-(4), we obtain
\begin{equation}
S_{f-g} = - \sum\limits_r p_r \log\, p_r \leq - \sum\limits_r
p_r \log \tilde p_r = -\sum\limits_r \tilde p_r \log \tilde p_r
= S\ ,
\label{fivefive}
\end{equation}
so that $S_{f-g}$ provides a lower bound for the entropy $S$. If the
initial condition and the coarse graining are related in such away that
$S$ is initially near its lower bound, then it will tend to increase for
a period of time.  That is the way the second law of thermodynamics
comes to hold.

In order to know what nearness to the lower bound means, we should examine
the upper bound on $S$. That upper bound is achieved when all
fine-grained alternatives have equal coarse-grained probabilities
$\tilde p_r$, corresponding in statistical mechanics to something like
``equilibrium'' or infinite temperature.  Each $\tilde p_r$ is then
equal to $N^{-1}$, where $N$ (assumed finite) is the number
of fine-grained alternatives, and the maximum entropy is thus
\begin{equation}
S_{\rm max} = \log\, N\ .
\label{fivesix}
\end{equation}

The simplest example of coarse graining utilizes a grouping of the
fine-grained alternatives $\{r\}$ into exhaustive and mutually exclusive
classes $\{\alpha\}$, where a class $\alpha$ contains  $N_\alpha$
elements and has lumped probability
\begin{equation}
p_\alpha \equiv \sum\limits_{r\in\alpha} p_r\ .
\label{fiveseven}
\end{equation}
Of course we have
\begin{equation}
\sum\limits_\alpha N_\alpha = N, \quad \sum\limits_\alpha
p_\alpha =  1\ .
\label{fiveeight}
\end{equation}
The coarse-grained probabilities $\tilde p_r$ in this example are the
class averages
\begin{equation}
\tilde p_r= p_\alpha/N_\alpha\ , \quad r\in\alpha\ ,
\label{fivenine}
\end{equation}
and they clearly have the properties (\ref{fivethree}). The entropy
comes out to be
\begin{equation}
S = -\sum\limits_\alpha p_\alpha \log \, p_\alpha +
\sum\limits_\alpha p_\alpha \log N_\alpha\ ,
\label{fiveten}
\end{equation}
where the second term contains the familiar logarithm of the number of
fine-grained alternatives (or microstates) in a
coarse-grained alternative (or macrostate), averaged over all the
coarse-grained alternatives.

Besides entropy, it is useful to introduce the concept of
algorithmic information content (AIC) as defined some thirty years ago (at the time of writing)
by Kolmogorov, Chaitin, and Solomonoff (all working
independently).\footnote{For a discussion of
the original papers see \cite{LV93}.}
For a string of bits
$s$
and a particular universal computer $U$, the AIC of the string,
written $K_U(s)$, is the length of the shortest program that
will cause $U$ to print out the string and then halt.
The string can be used as the description of some entity $e$, down
to a given level of detail, in a given language, assuming a given amount
of knowledge and understanding of the world, encoded in a
given way into bits \cite{GM95}. The AIC of the string can then be regarded as
$K_U(e)$, the AIC of the entity so described.

We now discuss a way of approaching classicality that utilizes AIC
as well as entropy. Some authors have tried to identify AIC in a
straightforward way with complexity, and in fact AIC is often called
algorithmic complexity. However, AIC is greatest for a ``random'' string
of bits with no regularity and that hardly corresponds to what is
usually meant by complexity in ordinary parlance or in scientific
discourse. To illustrate the connections among AIC, entropy or information,
and an effective notion of complexity,
take the ensemble $\widetilde E$ consisting of a set of fine-grained
alternatives $\{r\}$ together with their coarse-grained probabilities
$\widetilde p_r$.
We can then consider both $K_U(\widetilde E)$, the AIC of the ensemble, and
$K_U(r | \widetilde E)$, which is the AIC of a particular alternative $r$ given
the ensemble. For the latter we have the well known inequality (see, for
example \cite{Ben82}):

\begin{equation}
\sum\limits_r \widetilde p_r K_U (r |\widetilde E) \geq - \sum_r
\widetilde p_r\log \widetilde p_r=S\ .
\label{fiveeleven}
\end{equation}
Moreover, it has been shown by  R. Schack \cite{Schpp}
that, for any
$U$, a slight modification $U\to U^\prime$ permits
$K_{U^\prime}(r|\widetilde E)$ to be bounded on both
sides as follows:
\begin{equation}
S+ 1 \geq \sum\limits_r \widetilde p_r K_{U^\prime} (r|\widetilde E)
\geq S,
\label{fivetwelve}
\end{equation}
so that we have
\begin{equation}
\sum\limits_r \widetilde p_r K_{U^\prime} (r|\widetilde E)\approx
S.
\label{fivethirteen}
\end{equation}
(Previous upper bounds had ${\cal O}(1)$ in place of 1, but there was
nothing to prevent ${\cal O}(1)$ from being millions or trillions of
bits!)

Looking at the entropy $S$
as a close approximation to $\sum_r
\widetilde p_r K_{U^\prime} (r | \widetilde E)$, we see that it
is natural to complete it by adding to it the quantity
$K_{U^\prime}(\widetilde E)$ --- the AIC of the {\it ensemble}
with respect to the same
universal computer $U^\prime$. This last quantity can be
connected with the idea of effective complexity --- the length of  the most
concise description of the perceived regularities of an entity $e$.
Any particular set of regularities can be expressed by describing $e$ as
a member of an ensemble $\widetilde E$ of possible entities sharing
those regularities.
Then $K_{U^\prime}(\widetilde E) $ may be identified
with the effective complexity of $e$  or of the ensemble $\widetilde E$
\cite{GL96,GM95}. Adding this effective complexity to $S$, we have:
\begin{equation}
\Sigma \equiv K_{U^\prime }(\widetilde E) +S \ .
\label{fivethirteena}
\end{equation}
This sum of the the effective complexity and the entropy (or Shannon
information) may be labeled either ``augmented entropy'' or ``total
information" If the coarse graining is the simple one obtained by
partitioning the set of fine-grained alternatives $\{r\}$ into classes
$\{\alpha\}$ with cardinal numbers $N_\alpha$, then the total
information becomes
\begin{equation}
\Sigma =  K_{U^\prime}(\widetilde E) -
\sum\limits_\alpha p_\alpha \log
p_\alpha + \sum\limits_\alpha p_\alpha \log N_\alpha\ .
\label{fivefourteen}
\end{equation}

In \eqref{fivethirteena}, the first term becomes smaller as the set
of perceived regularities becomes simpler, while the second term becomes
smaller as the spread of possible entities sharing those regularities is
reduced. Minimizing $\Sigma$ corresponds to optimizing the choice of
regularities and the resulting effective complexity thereby  becomes
less subjective. Thus,
the total information or augmented entropy is useful in a wide variety
of contexts \cite{GL96,GM95}.  We apply it here to sets of  alternative
 decohering coarse-grained histories in quantum mechanics.

The general idea of augmenting entropy with a term referring to
algorithmic information  content was proposed
in a different context by Zurek \cite{Zur89}. However, as far as we
know, the emphasis on the utility of the quantity $\Sigma$ in
(\ref{fivethirteena}) and (\ref{fivefourteen}) is new.
We discussed the general idea of an entropy for histories
in \cite{GH90a}. Earlier, Lloyd and
Pagels \cite{LP88} introduced a quantity they called thermodynamic depth,
applicable to alternative coarse-grained classical histories $\alpha$.
They defined it as
\begin{equation}
D = \sum\limits_\alpha p_\alpha \log (p_\alpha/q_\alpha)\ ,
\label{fivefifteen}
\end{equation}
where $q_\alpha$ is an ``equilibrium probability'', which in our
notation would be $N_\alpha/N$ for the simple coarse graining we have
discussed.  We clearly have
\begin{equation}
D= \log N + \sum\limits_\alpha p_\alpha \log p_\alpha -
\sum\limits_\alpha p_\alpha \log N_\alpha
\label{fivesixteen}
\end{equation}
or
\begin{equation}
D=S_{\rm max} - S
\label{fiveseventeen}
\end{equation}
for the set of alternative coarse-grained histories. We see that
thermodynamic depth is intimately related to the notion of an entropy
for histories.

In applying augmented entropy to sets of coarse-grained histories in
quantum mechanics, one must take into account that there are
infinitely many different sets of fine-grained histories and that these
sets do not
generally have probabilities because they fail to decohere.
The quantities $N_\alpha$ may therefore conceivably depart from their
obvious definition as the numbers of fine-grained histories in the
coarse-grained classes $\{\alpha\}$. In fact, there may be
some latitude in the precise definition of the
complexity and entropy terms in the total information
(\ref{fivethirteena}). For example, one could consider instead of
$\widetilde E$ an ensemble
$\hat E$ consisting of the coarse-grained histories
$\alpha=(\alpha_1,\alpha_2,\cdots,\alpha_n)$, their probabilities
$p_\alpha$, and the numbers $N_\alpha$. A more general definition
of the entropy $S$ of histories  may help to define these numbers.
The generalized Jaynes construction for coarse-grained histories
 provides one framework for doing this \cite{GH90a}.
In the most general situation, such a construction defines the
entropy $S$ as the maximum of
$-Tr(\tilde \rho \log \tilde \rho)$ over all density matrices $\tilde \rho$
that
preserve the decoherence and probabilities of a given ensemble $E$ of
coarse-grained histories.  Other Jaynes-like constructions may also be useful,
for example ones that
define entropy by proceeding step by step through the histories

In any case,
our augmented entropy in (15) for coarse-grained decohering histories
in quantum mechanics is a negative measure of
classicality: the smaller the quantity, the closer the set of
alternative histories is to a quasiclassical realm.
Reducing the first term in (15) favors making the description of the
sequences of projections simple in terms of the field variables of the
theory and the Hamiltonian $H$.
It favors sets of
projections at different times that are related to one another by time
translations, as are many sequences of projections on quasiclassical
alternatives
at different times in the our quasiclassical realm.

Reducing the second term favors more nearly deterministic situations in
which the spread of probabilities is small. Approximate determinism
is, of course, a property of a quasiclassical realm. Reducing the last term
corresponds roughly to approaching ``maximality'', allowing the finest
graining that still permits decoherence and nearly classical behavior.
A quasiclassical realm must be maximal in order for it to be a feature
exhibited by the quantum state  and Hamiltonian and not a matter of
choice by an observer.

Any proposed measure of closeness to a quasiclassical realm must be
tested by searching for pathological cases of alternative decohering
histories that make the quantity small without resembling
quasiclassical realm of everyday experience.
The worst pathology occurs for a set of histories
in which the $P$'s at every time are projections on $|\Psi\rangle$ and
on states orthogonal to $|\Psi\rangle$. We see that in this pathological
case the description of the histories and their probabilities is simple
because the description of the initial state is simple, so that
$K_{U'}(\widetilde E)$ is small. The term $-\sum_\alpha  p_\alpha {\rm log}
p_\alpha$
is zero
and the third term is also zero since the only $\alpha$ with
$p_\alpha \neq  0$ corresponds to projecting onto the pure state
$|\Psi\rangle$, so that $N_\alpha$ is one and ${\rm log}N_\alpha$
vanishes.

Evidently the smallness of $\Sigma$ is not by itself a
sufficient criterion for characterizing a quasiclassical realm.
Further criteria can be introduced if we require that
quasiclassical
realms be  decohering with suitable restrictions on the operators from which the future histories are
constructed. Requiring  decherence   ensures probabilities consistent with the usual rules of probability theory.

The sets  must be restricted so as to rule out pathologies
such as discussed above.
Presumably they must all belong to a huge set with certain
straightforward properties.
Those properties might be
connected with locality, since quantum field theory is perfectly local.
(Even superstring theory is local --- although the string is an extended
object, interaction among strings is always local in spacetime.)
 It would be in this way that  decoherence enters a definition
of classicality. (Some  some information theoretic measures not unrelated to the one for classicality discussed
here can be found in \cite{anas99,GL96,Zur89}.)

The general description of the set of histories of a quasiclassical realm is intuitively simple in several respects. 
They are coarse-grained by a few quasiclassical variables like the hydrodynamic variables discussed in Section \ref{QCR}. Obeying  deterministic laws the description of individual histories can be reduced to initial conditions.
That is why the measure of classicality introduced in Section \ref{measure} consists of two parts.  An entropy term to favor determinism and an AIC term to favor siimplicity. Classicality requires both.

\section{Using  the measure to define  `quasiclassical realm,`}
\label{qcr}
Given the measure  of classicalty defined  in Section \ref{measure} , a quasiclassical could be characterized in quantum
mechanics as a realm that minimizes the augmented
entropy given by (15) subject to further suitable conditions. Quasiclassical realms so defined would be an
emergent feature of $H$, $|\Psi\rangle$, and quantum mechanics --- a
feature of the universe independent of human choice. In principle,
given $H$ and $|\Psi\rangle$, we could compute the quasiclassical realm
that these theories exhibit. We could then investigate the important
question of whether our  quasiclassical realm is essentially unique
or whether the quantum mechanics of the universe exhibits essentially
 inequivalent other quasiclassical
realms. Either conclusion would be of central importance for
understanding quantum mechanics.

\section{Our Place in Our  Quasiclassical Realm}
\label{QCR}
As observers of the universe we are physical systems within it. Both individually and collectively we are information gathering and utilizing systems (IGUSes)  described in quasiclassical terms and obeying classical laws. As physical systems we are  part of the our  classical realm defined  coarse-grained histories of hydrodynamic variables. these are integrals of conserved quantities like energy, momentum, and conserved quantities  like baryon number over suitably sized  volumes.    
The evidence of our  observations of the Universe  suggests such a quasiclasiscal realm  extends approximately over patch of classical spacetime the size of a Hubble volume\footnote{By Hubble volume we mean a region of space whose size is roughly the time from the big-bang to the present.}.  As IGUSes we  function by exploiting the regularities such a quasiclassical realm exhibits \cite{har16a} .

\section{Other Quasiclassical Realms}
\label{OQCR}
It is possible that our Universe exhibits a multiverse of quasiclassical realms \cite{multi} ---- more than one quasiclassical realm in different locations in space and/or  more than one kind of quasiclassical realm,. The classical spacetimes inside bubbles of true vacuum nucleated by the decay of a false vacuum provide a simple examples 
\cite{HH16}. 

These different quasiclassical  realms could be based on variables different from the hydrodynamic variables defined above. They could have different low energy physics, and different levels of classicality as  defined by the measure in Section II. 

Could  these different quasiclassical realms have different kinds and numbers of IGUSs.?Could we communicate with them if they did?  Such questions are beyond our power to answer now. But it is possible to imagine that they could be addressed in the future both theoretically and experimentally. The classicality measure developed in this paper would help.

\section{Modeling Qiuasiclassical Realms}
\label{modeling} 
The measure of classicality defined in Section \ref{measure} could be better understood by calculating its value in simple model closed systems.   

One the simplest models is the one-dimensional quantum harmonic chain described in \cite{BruHar99}. A closed quantum system consists of a one-dimensional chain of particles of equal mass $m$ interacting with each other by nearest neighbor harmonic potentials.  Fine-grained histories are specified by giving the position of each particle as a function of time. 

Another example are the realms defined by minisuperspace models in quantum cosmology eg.\cite{HHH08,Houches}  in the context of semiclassical quantum gravity. Imagine calculating the measure  of classicality assuming that all the histories are homogeneous and isotropic. 
Does the classicality go down or up if histories with small quantum fluctuations are included in realm?  Does classicality go down or up if histories with small fluctuations are included in the realm?

It would be a major research effort to work out either of these models but our understanding of what is classical would be improved thereby.

\section{Acknowledgments}. The junior author acknowledges the importance of the many conversations with his collaborator.  The work  of JH was supported  in part by US National Science Foundation grants  PHY-15004541 and  in the past by PHY-12005500 and PHY15-04541.  Acknowledgments to the paper Strong Decoherence \cite{GH95}  from  which Section \ref{measure} was taken can be found there.


\end{document}